\documentclass[twocolumn]{jpsj2}

\title{
Specific Heat Study of Magnetic and Superconducting Transitions 
in CePt$_{3}$Si 
}

\author{
Gaku \textsc{Motoyama}, Katsuhiro \textsc{Maeda}, and 
Yasukage \textsc{Oda} \\
}

\inst{
Graduate School of Material Science, University of Hyogo, \\
Kamigori-cho, Ako-gun, Hyogo 678-1297, Japan. \\
}

\abst{
Measurements of specific heat between 80 mK to 4 K and 
electrical resistivity between 80 mK to 10 K 
were carried out for polycrystalline CePt$_{3}$Si samples 
cut into small pieces (typically $\sim $10 mg). 
In the specific heat measurements, 
we observed an antiferromagnetic transition jump 
at $T_{\rm N}$ = 2.2 K for all the samples, 
while the heights have large variations. 
As regards superconductivity, 
we observed two distinct transition jumps at 
$T_{\rm c}^{\rm l}$ $\sim$ 0.45 K and $T_{\rm c}^{\rm h}$ $\sim$ 0.75 K, 
which were the same for all the samples. 
From the measurements of specific heat and resistivity, 
systematic relations were found between antiferromagnetic 
and superconducting transitions. 
We conclude that antiferromagnetism, 
whose transition temperature is 2.2 K, 
coexists with superconductivity, 
whose transition temperature is $T_{\rm c}^{\rm l}$. 
In this sample, 
residual electronic specific heat coefficient 
in the superconducting state $\gamma_{\rm s}$ was quite small, 
and specific heat divided by temperature below $T_{\rm c}^{\rm l}$ 
decreased almost linearly with decreasing temperature. 
In order to reveal the characteristic properties of the 
magnetism and superconductivity of the CePt$_{3}$Si system, 
it is important to study the two superconducting phases 
with $T_{\rm c}^{\rm l}$ and $T_{\rm c}^{\rm h}$, 
respectively. 
}

\kword{CePt$_{3}$Si, non-centrosymmetry, antiferromagnetism, 
heavy fermion superconductor, specific heat measurement
}

\begin{document}
\maketitle

\section{Introduction}

Bauer {\it et al.} reported that CePt$_{3}$Si exhibits 
an antiferromagnetic (AFM) order at $T_{\rm N}$ = 2.2 K 
and superconducting (SC) transition at $T_{\rm c}$ = 0.75 K, 
and that normal state electronic specific heat coefficient $\gamma_{\rm n}$ is 
approx 0.39 J/(K$^{2}\cdot $mol)\cite{EBauer}. 
This compound is a heavy fermion superconductor having a characteristic 
crystal structure that lacks inversion symmetry (space group $P4mm$). 
Therefore, 
the superconductivity of CePt$_{3}$Si exists in a particular environment 
compared with that of a conventional superconductor. 
Previous theoretical studies have shown that 
a non-centrosymmetric heavy fermion has several possible states 
for realizing unconventional 
superconductivity\cite{PAFrig, SFuji, NHaya, HTana, YYana}.

Many experimental studies of superconductivity have been carried out. 
Previous studies of specific heat by Bauer {\it et al.} and Takeuchi {\it et al.} 
have shown marked contrasts 
between polycrystalline and single crystal samples\cite{EBauer, TTake}. 
The former showed a small AFM transition jump at $T_{\rm N}$ = 2.2 K and 
an SC transition jump at $T_{\rm c}$ = 0.75 K for a polycrystalline sample. 
The latter showed a large AFM transition jump at the same $T_{\rm N}$ and 
an SC transition jump at different $T_{\rm c}$ of 0.46 K for a single crystal sample. 
In the single crystal sample, 
the SC jump was sharp and large, 
and the residual electronic specific heat coefficient in the SC state, 
$\gamma_{\rm s}$, was small, 
however, its $T_{\rm c}$ was low compared with that of the polycrystalline sample. 
In addition, a double anomaly of the SC state 
in the specific heat measurement was observed 
by Scheidt {\it et al.}\cite{ESchei} 
They suggest that 
it was a signalling two consecutive phase transitions. 
On the other hand, 
Nakatsuji {\it et al.} showed that 
the Meissner effect of SC started increasing from $\sim $ 0.8 K 
and the rate of increase changed below $\sim $ 0.5 K\cite{KNaka}. 
They suggest that the SC domain has a volume fraction. 
The pressure dependence of the Meissner effect and $T_{\rm c}$ seemed to indicate 
that the volume fraction was due to some inhomogeneous property that 
leads to a spatial variation of local pressure in the sample\cite{YAoki}.

The temperature $T$ dependence of specific heat divided by temperature 
($C/T$) in the SC region preferred a linear $T$ dependence 
over a $T^{-1}$exp$(-\Delta /k_{B}T)$ dependence\cite{TTake}. 
In addition, the $T$ dependence of thermal conductivity in the $T$ 
range of 40 mK to 0.2 K was well fitted by a linear function of $T$\cite{KIzawa}. 
These results indicate the presence of line nodes in the SC energy gap. 
On the other hand, 
the $T$ dependence of nuclear spin-lattice relaxation rate $1/T_{1}$($T$) 
did not simply follow an exponential law or a $T^{3}$-power law. 
The plot of $1/T_{1}T$($T$) showed a coherence Hebel-Slichter peak at $T_{\rm c}$, 
indicating a full-gap state without nodes\cite{MYogi}. 
Another NMR measurement indicated that 
the plot of $1/T_{1}$($T$) showed 
no obvious Hebel-Slichter peak and a drastically decreasing 
($\propto T^{3}\sim T^{5}$)\cite{KUeda1, KUeda2}. 
Therefore, CePt$_{3}$Si is expected to be an unconventional superconductor. 
In addition, 
the pressure $P$ phase diagram of $T_{\rm c}$ and $T_{\rm N}$ for this system 
was unusual compared with that of the previous magnetic 
superconductor\cite{TYasu, NTate, Bauer2}. 
Although $T_{\rm c}$ and $T_{\rm N}$ decreased with increasing $P$, 
SC still existed even after AFM disappeared. 
The decreasing rate of $T_{\rm c}$ slowed down only at 
around $P_{\rm c}$, at which AFM disappeared. 
The pressure corresponding to the maximum $T_{\rm c}$ in this system 
was not $P_{\rm c}$. 
Some heavy fermion magnetic superconductors show a dome structure 
for the pressure dependence of $T_{\rm c}$ at the critical point $P_{\rm c}$.

\section{Experimental}

Polycrystalline CePt$_{3}$Si and Ce$_{1.01}$Pt$_{3}$Si samples 
were synthesized by arc-melting Ce of 99.9 \% (3N) purity, Pt of 3N5 purity, 
and Si of 6N purity, using a laboratory-made furnace. 
The synthesized melt became solidified by quenching on Cu-hearth 
in Ar atmosphere of 6N purity. 
The chemical compositions of our samples were determined 
from those of the starting materials. 
The weight loss of the constituent materials was negligible 
during the preparation. 
An ingot sample (1$\sim $2 g) was cut into two lumps, 
and one lump was heat-treated. 
Heat treatment for annealing was carried out under well-controlled conditions: 
the temperature was maintained at 950$^\circ $C for one week 
and lowered to room temperature over three days. 
We labeled heat-treated and non-heat-treated samples as 
"annealed" and "as-cast", respectively. 
Then, each lump was cut into small pieces ($\sim$ 10 mg) 
for measurement. 
We prepared three CePt$_{3}$Si as-cast (\#1, 2 and 3), 
two Ce$_{1.01}$Pt$_{3}$Si as-cast (\#4 and 5), 
and their annealed samples (\#1-a, \#2-a and so on) 
to investigate sample dependence\cite{GMoto1, GMoto2}. 
Moreover, we conducted measurements using different pieces 
from the same batch (\#2-a-1, \#2-a-2, and so on).

Temperature dependence of specific heat was measured using 
the adiabatic heat pulse method between $\sim $80 mK to 4 K. 
Electrical resistivity was measured using 
the conventional dc four-terminal method down to $\sim $80 mK 
using the same piece as that used in specific heat measurement. 
Measurements were carried out using a laboratory-made dilution 
refrigerator.

\section{Results and Discussion}

\begin{figure}
\begin{center}
\includegraphics[width=8.5cm]{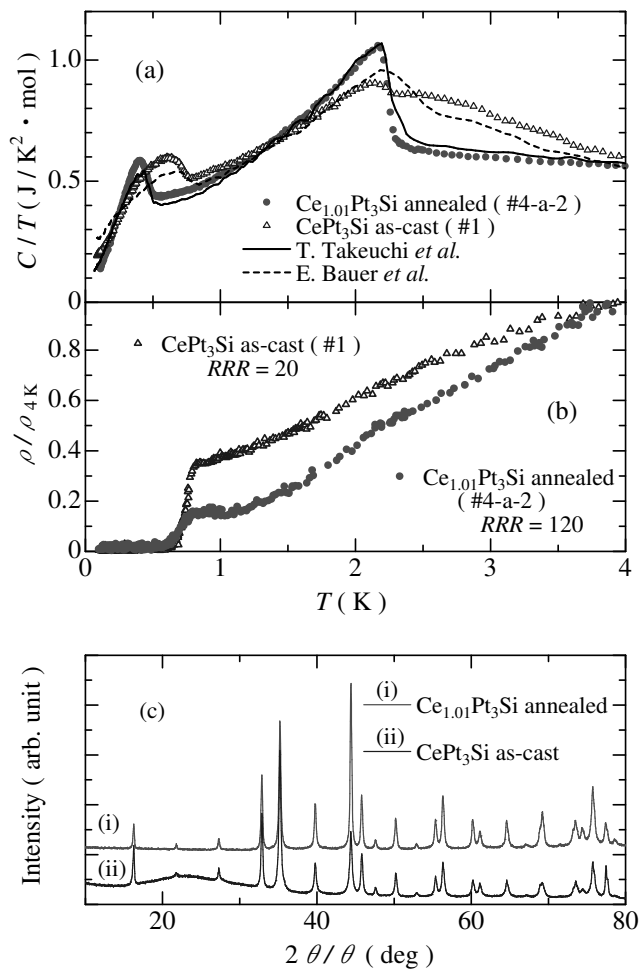}
\end{center}
\caption{
(a) Temperature dependence of specific heat divided by temperature, $C/T$, for 
CePt$_{3}$Si as-cast (\#1) and Ce$_{1.01}$Pt$_{3}$Si annealed (\#4-a-2). 
The solid line shows data obtained by Takeuchi {\it et al.} (ref. 7), 
and the broken line shows data of Bauer {\it et al.} (ref. 1). 
(b) Temperature dependence of electrical resistivity, which was measured 
using the same pieces as those in Fig. 1(a). 
CePt$_{3}$Si as-cast (\#1) had a small residual resistivity ratio ($RRR$) of 20; 
Ce$_{1.01}$Pt$_{3}$Si annealed (\#4-a-2) had a large $RRR$ of 120. 
(c) X-ray diffraction patterns of samples 
from the same batch as those shown in Figs. 1(a) and 1(b). 
Pattern (ii) contained a larger background effect than pattern (i). 
} 
\label{f1}
\end{figure}

Figure 1(a) shows the $T$ dependence of specific heat divided by $T$ ($C/T$) 
of the CePt$_{3}$Si as-cast (\#1) and Ce$_{1.01}$Pt$_{3}$Si annealed (\#4-a-2) samples. 
They showed quite different characteristics despite having the same polycrystalline 
CePt$_{3}$Si system. 
The $C/T(T)$ of the Ce$_{1.01}$Pt$_{3}$Si annealed showed a 
distinct AFM transition with a large jump at $T_{\rm N}$ (= 2.2 K) 
and SC transition with a sharp jump at low $T_{\rm c}$ (= 0.45 K). 
The residual $\gamma_{\rm s}$ extrapolated to 0 K was almost zero. 
These results were similar to those reported by Takeuchi {\it et al.} 
for their single crystal, as shown by the solid line. 
On the other hand, the $C/T(T)$ of CePt$_{3}$Si as-cast exhibited 
a very small jump of AFM transition and a jump of SC transition 
appearing at high $T_{\rm c}$ (=0.75 K) 
compared with that of Ce$_{1.01}$Pt$_{3}$Si annealed. 
The AFM transition of this sample had not only a small jump 
but also a broad tail above $T_{\rm N}$, from 2.2 K to $\sim$ 4.0 K. 
$C/T(T)$ clearly increased at the onset of $T_{\rm c}$ with decreasing $T$, 
but the peak broadened. 
These behaviors were similar to that observed by Bauer {\it et al.} 
for their polycrystalline sample, as shown by the broken line. 
Figure 1(b) shows the temperature dependence of the electrical resistivity ($\rho $) 
of the CePt$_{3}$Si as-cast and Ce$_{1.01}$Pt$_{3}$Si annealed samples. 
Measurements were carried out using very small pieces. 
Because the absolute value might include some ambiguity, 
$\rho/\rho_{4 K}(T)$ values are presented. 
The residual resistivity ratio ($RRR$) of CePt$_{3}$Si as-cast was 20 
and that of Ce$_{1.01}$Pt$_{3}$Si annealed was 120. 
We confirmed reproducibility by some measurements. 
All of the measured Ce$_{1.01}$Pt$_{3}$Si annealed including different batches 
had $RRR$ exceeding 100. 
These were remarkably large, 
but other measured samples indicated $RRR$ $\sim$ 20. 
A kink of $\rho/\rho_{4 K}(T)$ was observed 
at $T_{\rm N}$=2.2 K for both samples. 
The $\rho/\rho_{4 K}(T)$ of the Ce$_{1.01}$Pt$_{3}$Si annealed showed 
a clear kink at $T_{\rm N}$ 
and decreased rapidly below $T_{\rm N}$ with decreasing temperature. 
The decrease plateaued immediately just above $T_{\rm c}$. 
On the other hand, 
the kink of the $\rho/\rho_{4 K}(T)$ of the CePt$_{3}$Si as-cast broadened, 
and the decrease of $\rho/\rho_{4 K}(T)$ continued to $T_{\rm c}$. 
These results of $C/T(T)$ and $\rho/\rho_{4 K}(T)$ indicate that 
Ce$_{1.01}$Pt$_{3}$Si annealed has a more regular AFM ordering 
(which is a long-range ordering with a narrow $T_{\rm N}$ at 2.2 K) and a 
more regular lattice (which is an ideal CePt$_{3}$Si lattice, that is a 
non-centrosymmetric lattice) than CePt$_{3}$Si as-cast. 
Figure 1(c) shows the X-ray diffraction patterns. 
No extra-phase was observed in the X-ray diffraction patterns of both samples. 
There was no difference in the accuracy of measurement 
between the lattice constants of the two samples. 
The results in Fig. 1 indicate that 
1\% variations in Ce-concentration and heat treatment yield 
small structural changes 
that strongly affect SC and AFM but not powder diffraction patterns. 
In our speculation, these structural changes concern the non-centrosymmetric structure, 
which is an ordering of Pt and Si atoms 
occupying the two 1(a)-sites of the $P4mm$ structure. 
Because, it is considered that non-centrosymmetricity is important 
for both SC of this system and AFM whose magnetic structure 
consists of ferromagnetic c-planes stacked antiferromagnetically 
along the c-axis\cite{NMetok}. 
Therefore, although Ce$_{1.01}$Pt$_{3}$Si annealed has 1\% opening Pt and Si sites, 
it might have two well-ordered 1(a)-sites of $P4mm$ 
occupied by Pt and Si atoms. 
Then, the opening sites might be available for removing and ordering Pt and Si atoms 
when a sample is heat-treated. 
Conversely, although CePt$_{3}$Si as-cast has a stoichiometric composition, 
it might have some disorders of Pt and Si at the two 1(a)-sites, 
because there is a quenching process in the preparation of polycrystalline samples.

\begin{figure}
\begin{center}
\includegraphics[width=8.5cm]{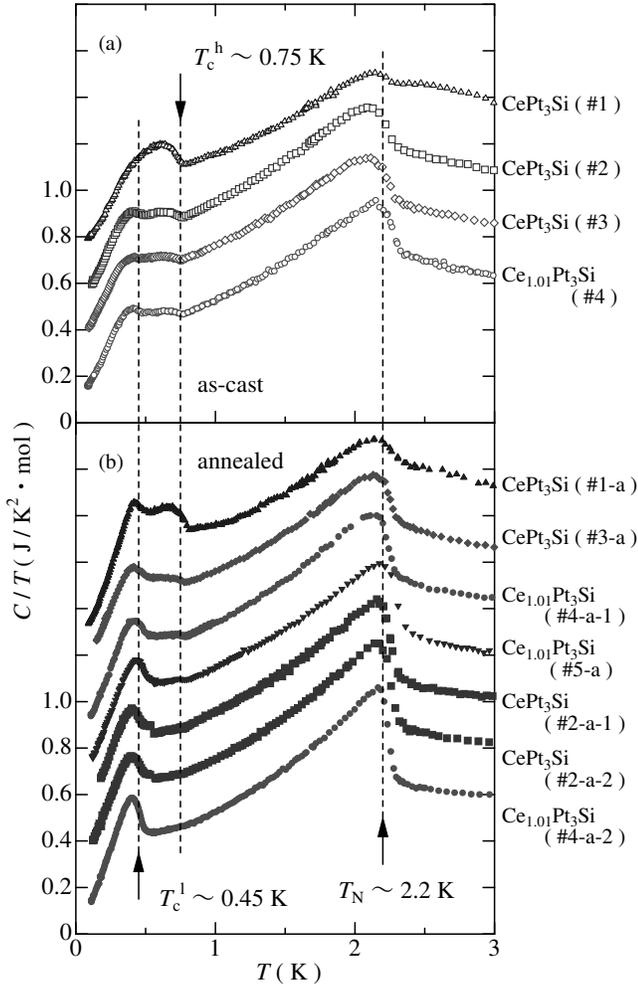}
\end{center}
\caption{
$T$ dependences of $C/T$ for 
CePt$_{3}$Si and Ce$_{1.01}$Pt$_{3}$Si. 
Experimental data were shifted upward by 0.2 J/(K$^{2}\cdot $mol) intervals. 
Data with open (closed) symbols in Figs. 2(a) and 2(b) were respectively 
measured using as-cast and annealed samples. 
$T_{\rm N}$ is the antiferromagnetic ordering temperature, and 
bulk superconductivity occurs below $T_{\rm c}^{\rm l}$ and/or $T_{\rm c}^{\rm h}$. 
Each sample was denoted as \#(sample number)-('a': annealed)-(piece number). 
} 
\label{f2}
\end{figure}

Figures 2(a) and 2(b) show the $C/T(T)$ of various samples 
in order to consider sample dependence in detail. 
Fig. 2(a) shows the results for the as-cast samples 
and Fig. 2(b) shows those for the annealed samples. 
In Figs. 2(a) and 2(b), 
the data were arranged according to the decrease in the height 
of the specific heat jump in AFM transition from bottom to top. 
First, we note both AFM and SC transitions, respectively. 
The height of the jump in AFM transition, $\Delta C/T(T_{\rm N})$, decreased gradually 
without changing $T_{\rm N}$, 
and the broad tail above $T_{\rm N}$ enlarged gradually from the bottom to top data. 
This relation is plotted in Fig. 3(b). 
We observed two peaks for SC transition 
in both Figs. 2(a) and 2(b). 
We define $T_{\rm c}^{\rm l}$ ($\sim$ 0.45 K) as the temperature 
of the specific heat peak at lower 
and $T_{\rm c}^{\rm h}$ ($\sim$ 0.75 K) as the onset temperature of 
the specific heat peak at higher. 
$T_{\rm c}^{\rm l}$ and $T_{\rm c}^{\rm h}$ 
were almost constant for all the samples. 
However, the heights of the specific heat jump at $T_{\rm c}^{\rm l}$ and 
$T_{\rm c}^{\rm h}$, $\Delta C/T(T_{\rm c}^{\rm l})$ and 
$\Delta C/T(T_{\rm c}^{\rm h})$, differed for each sample. 
As $\Delta C/T(T_{\rm c}^{\rm l})$ increased, 
$\Delta C/T(T_{\rm c}^{\rm h})$ decreased. 
These results indicate that $T_{\rm c}^{\rm l}$ does not move to $T_{\rm c}^{\rm h}$ 
and that the SC at $T_{\rm c}^{\rm l}$ and $T_{\rm c}^{\rm h}$ 
compete against each other. 
CePt$_{3}$Si as-cast (\#1) has only one large peak at $T_{\rm c}^{\rm h}$. 
However, this peak might include some components of $\Delta C/T(T_{\rm c}^{\rm l})$, 
because it is broadened from $T_{\rm c}^{\rm h}$ to $T_{\rm c}^{\rm l}$. 
Unfortunately, we were unable to prepare a sample that shows only a sharp jump 
at $T_{\rm c}^{\rm h}$ and a small residual $\gamma_{\rm s}$. 
Next, the relations between the two SC transition jumps, 
$\Delta C/T(T_{\rm c}^{\rm l})$ and $\Delta C/T(T_{\rm c}^{\rm h})$, 
and the AFM transition jump, $\Delta C/T(T_{\rm N})$, should be noted. 
In Figs. 2(a) and 2(b), 
$\Delta C/T(T_{\rm c}^{\rm l})$ increased from the top to bottom data, 
that is, as $\Delta C/T(T_{\rm N})$ increased, 
$\Delta C/T(T_{\rm c}^{\rm l})$ increased as well. 
The most typical example of this case is Ce$_{1.01}$Pt$_{3}$Si annealed (\#4-a-2). 
$\Delta C/T(T_{\rm c}^{\rm h})$ increased from the bottom to top data. 
It should be noted that $\Delta C/T(T_{\rm N})$ and $\Delta C/T(T_{\rm c}^{\rm l})$ 
almost vanished in the sample with the largest $\Delta C/T(T_{\rm c}^{\rm h})$. 
These data of correlation between $\Delta C/T(T_{\rm c}^{\rm h})$, 
$\Delta C/T(T_{\rm c}^{\rm l})$ and $\Delta C/T(T_{\rm N})$ are plotted in Fig. 3(a). 
These relations are described later. 
Next, we compare the annealed samples in Fig. 2(b) 
with the as-cast samples in Fig. 2(a). 
The $\Delta C/T(T_{\rm N})$ and $\Delta C/T(T_{\rm c}^{\rm l})$ of the annealed samples 
were almost larger than those of the as-cast samples, 
while the $\gamma_{\rm s}$ of the annealed samples 
were smaller than those of the as-cast samples. 
These results are shown in Figs. 3(a) - 3(c) as open and closed symbols for 
as-cast and annealed, respectively.

\begin{figure}
\begin{center}
\includegraphics[width=8.5cm]{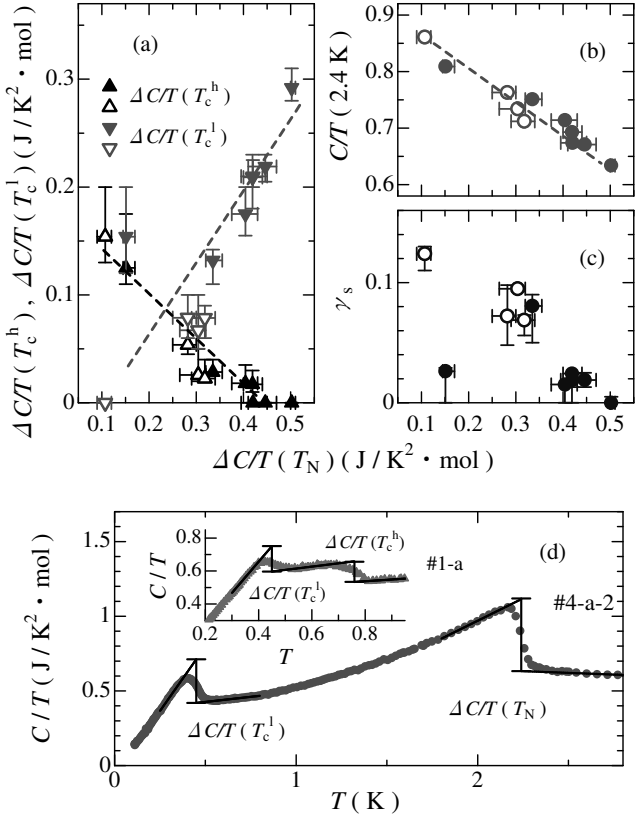}
\end{center}
\caption{
Relations of 
$\Delta C/T(T_{\rm c}^{\rm l})$, $\Delta C/T(T_{\rm c}^{\rm h})$, $C/T$(2.4 K) 
and $\gamma_{\rm s}$ versus $\Delta C/T(T_{\rm N})$ are plotted in Fig. 3(a) 
- 3(c), respectively. 
The broken lines are guides to the eye. 
Open and closed symbols indicate the results of as-cast and annealed, respectively. 
$\Delta C/T(T_{\rm c}^{\rm l})$ ($\Delta C/T(T_{\rm c}^{\rm h})$, 
$\Delta C/T(T_{\rm N})$) was determined by a linear extrapolation of the $C/T(T)$ data 
below $T_{\rm c}^{\rm l}$ ($T_{\rm c}^{\rm h}$, $T_{\rm N}$) and 
above $T_{\rm c}^{\rm l}$ ($T_{\rm c}^{\rm h}$, $T_{\rm N}$) 
to $T_{\rm c}^{\rm l}$ ($T_{\rm c}^{\rm h}$, $T_{\rm N}$), as indicated in Fig. 3(d). 
$\gamma_{\rm s}$ was decided by linear extrapolation to 0 K. 
} 
\label{f3}
\end{figure}

The relations of 
$\Delta C/T(T_{\rm c}^{\rm h})$, $\Delta C/T(T_{\rm c}^{\rm l})$, $C/T$(2.4 K) 
and $\gamma_{\rm s}$ versus $\Delta C/T(T_{\rm N})$ are plotted in Figs. 3(a) 
- 3(c), respectively. 
In Fig. 3(a), when $\Delta C/T(T_{\rm N})$ increases, $\Delta C/T(T_{\rm c}^{\rm h})$ 
decreases and $\Delta C/T(T_{\rm c}^{\rm l})$ increases. 
It is clear that SC at $T_{\rm c}^{\rm h}$ and 
$T_{\rm c}^{\rm l}$ compete against each other, and that 
SC at $T_{\rm c}^{\rm h}$ is on competitive relation with AFM at $T_{\rm N}$ 
but SC at $T_{\rm c}^{\rm l}$ is not. 
In Fig. 3(b), $C/T$(2.4 K), which reflects the broad tail above $T_{\rm N}$, 
increased with decreasing $\Delta C/T(T_{\rm N})$. 
This enhancement might have some relation 
with an increase in $\Delta C/T(T_{\rm c}^{\rm h})$. 
In Fig. 3(c), the relation between $\gamma_{\rm s}$ and $\Delta C/T(T_{\rm N})$ 
is not clear, but at least the $\gamma_{\rm s}$ of as-cast was large and 
the $\gamma_{\rm s}$ of annealed with a large $\Delta C/T(T_{\rm N})$ was small. 
$\Delta C/T(T_{\rm c}^{\rm l})$, $\Delta C/T(T_{\rm c}^{\rm h})$ and 
$\Delta C/T(T_{\rm N})$ were decided in accordance with Fig. 3(d). 
These absolute values have some ambiguities 
because of their broadness. 
However, it has no significant effects on their relations.

As mentioned above, the present experiment leads us to conclude 
that the CePt$_{3}$Si system 
is spatially separated into two superconducting regions, 
SC$^{\rm l}$ and SC$^{\rm h}$, 
whose transition temperatures are $T_{\rm c}^{\rm l}$ ($\sim$ 0.45 K) and 
$T_{\rm c}^{\rm h}$ ($\sim$ 0.75 K), respectively. 
SC$^{\rm l}$ develops in a more regular AFM ordering and a more regular lattice. 
Ce$_{1.01}$Pt$_{3}$Si annealed (\#4-a-2) is considered to have an almost single phase 
in which SC$^{\rm l}$ and AFM with $T_{\rm N}$ coexist. 
Because, the volume fraction of this sample exhibited SC$^{\rm l}$ 
with a particularly small residual $\gamma_{\rm s}$ and 
AFM with the most distinct and largest peak for this sample 
are probably bulk properties. 
On the other hand, SC$^{\rm h}$ does not seem to coexist with AFM 
having $T_{\rm N}$=2.2 K at least. 
In what kind of phase is this SC$^{\rm h}$ included? 
To answer this question, 
the broad tail gradually enlarging above $T_{\rm N}$ 
might give us some hints. 
We suggest that 
the region of SC$^{\rm h}$ is included in some magnetic phase 
which causes enhancement of the broad tail above $T_{\rm N}$ 
(for example, a heavy fermion non magnetic phase and 
an AFM phase with broad $T_{\rm N}$ from 2.2 to 4.0 K.\cite{WHige}) 
We suggested in our earlier discussion of Fig. 1 and in refs. 18 and 19 
that as-cast samples contain some defects, 
which are reduced in number by heat treatment and annealing. 
The annealed samples exhibit large $RRR$, 
a marked transition at a narrow $T_{\rm N}$ and 
a less ferromagnetically anomaly at 3.0 K. 
A sample having a perfectly regular structure 
would have perfect non-centrosymmetry. 
The presence of some defects will affect non-centrosymmetry 
and produce some partial disorders.
In particular, 
we consider Pt and Si atoms occupying the two 1(a)-sites of the $P4mm$ structure 
as important parts. 
From this viewpoint, 
there is a relation between the inhomogeneity of this system 
and the two volume fractions of SC$^{\rm l}$ and SC$^{\rm h}$. 
The as-cast sample that has some disorders in the non-centrosymmetric structure 
exhibits a large volume fraction of SC$^{\rm h}$, 
and the annealed sample with well-ordered non-centrosymmetricity 
exhibits a large volume fraction of SC$^{\rm l}$. 
It might be implied that 
SC$^{\rm h}$ develops in the centrosymmetric part and 
SC$^{\rm l}$ develops in the non-centrosymmetric part.

Here, we need to explain why the regular lattice 
has a low superconducting transition temperature, $T_{\rm c}^{\rm l}$. 
From the above sample characterization, 
we found that SC was affected by the AFM state 
and the disorder of the non-centrosymmetric structure. 
We suggest three scenarios to explain $T_{\rm c}^{\rm l}$. 
(i) One scenario is that SC$^{\rm h}$ exists in the 
non magnetic heavy fermion state 
in contrast to the coexistence of SC$^{\rm l}$ 
and AFM at $T_{\rm N}$ of 2.2 K. 
$T_{\rm c}^{\rm l}$ might be reduced by an internal magnetic field of AFM. 
In this case, the broad tail of $C/T(T)$ above $T_{\rm N}$ is due to 
a non magnetic heavy fermion state. 
(ii) The second scenario is that SC$^{\rm h}$ exists in another 
inhomogeneous AFM phase with a broad transition temperature 
from 2.2 K to 4.0 K, $T_{\rm N}^{\rm h}$. 
This inhomogeneous AFM phase transition is the cause of the broad tail. 
This scenario is consistent with the $P$-$T$ phase diagram. 
In the $P$-$T$ phase diagram, 
$T_{\rm c}$ and $T_{\rm N}$ decrease with increasing pressure. 
If effective pressure caused by the inhomogeneity of a lattice 
were reduced in the region of SC$^{\rm h}$, 
the enhancement to $T_{\rm c}^{\rm h}$ and the broad tail of $T_{\rm N}^{\rm h}$ 
could be explained\cite{YAoki}. 
A degree of the inhomogeneity causes a broad $T_{\rm c}^{\rm h}$ 
and a broad tail of $T_{\rm N}^{\rm h}$. 
(iii) The third scenario is that SC$^{\rm l}$ exists 
in a well-ordered non-centrosymmetric region 
and SC$^{\rm h}$ exists in a disordered region. 
Some previous theoretical works have shown that 
$T_{\rm c}$ is suppressed by enhancing 
antisymmetric spin-orbit coupling\cite{HTana, YYana}. 
In general, antisymmetric spin-orbit coupling is weakened 
by the disorder of a non-centrosymmetric structure. 
Therefore, $T_{\rm c}^{\rm l}$ was suppressed 
by enhancing antisymmetric spin-orbit coupling, 
and $T_{\rm c}^{\rm h}$ was hardly suppressed. 
These are simple illustrations. 
In fact, 
it might be described by combining the above scenarios and others.

\begin{figure}
\begin{center}
\includegraphics[width=8.0cm]{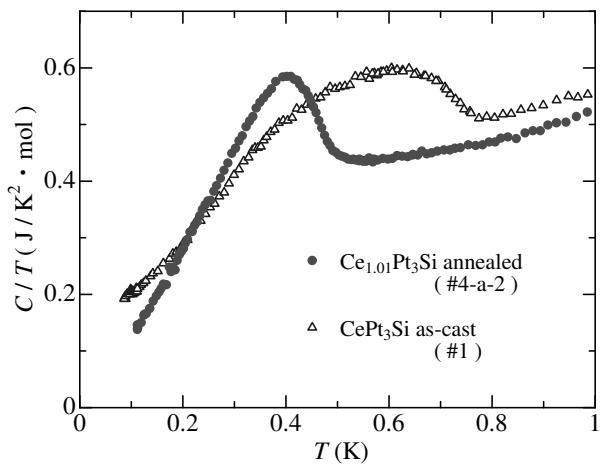}
\end{center}
\caption{
$T$ dependences of $C/T$ below 1 K for 
Ce$_{1.01}$Pt$_{3}$Si annealed and CePt$_{3}$Si as-cast. 
} 
\label{f4}
\end{figure}

Figure 4 shows the magnified $C/T(T)$ of the Ce$_{1.01}$Pt$_{3}$Si annealed (\#4-a-2) 
and CePt$_{3}$Si as-cast (\#1) samples below 1 K. 
The former exhibited the largest jump at $T_{\rm c}^{\rm l}$, 
while the latter exhibited the largest jump at $T_{\rm c}^{\rm h}$. 
As shown in the figure, the $C/T$ of sample \#4-a-2 decreased linearly to temperatures 
below $T_{\rm c}^{\rm l}$ and the extrapolated residual $\gamma_{\rm s}$ to 0 K 
was very small. 
In contrast, the $C/T$ of sample \#1 seemed to have a different temperature dependence, 
and the extrapolated residual $\gamma_{\rm s}$ to 0 K was finite. 
In order to reveal the $T$ dependence of $C/T$ below $T_{\rm c}^{\rm h}$, 
we need to prepare a sample with only a sharp jump at $T_{\rm c}^{\rm h}$ and 
a small residual $\gamma_{\rm s}$.

\section{Conclusions}

In conclusion, 
we observed an AFM transition jump in specific heat measurements 
of the polycrystalline CePt$_{3}$Si system. 
We observed $T_{\rm N}$ = 2.2 K for all the measured pieces, 
but $\Delta C/T(T_{\rm N})$ tended to vary among pieces. 
As $\Delta C/T(T_{\rm N})$ decreased, 
a broad tail above $T_{\rm N}$ enlarged gradually. 
We observed two SC transition jumps at 
$T_{\rm c}^{\rm l}$ and $T_{\rm c}^{\rm h}$, 
which showed no sample dependence. 
SC$^{\rm l}$ and SC$^{\rm h}$ volume fractions are considered to be 
spatially separated to each other in a piece. 
A larger $\Delta C/T(T_{\rm c}^{\rm l})$ appeared in a piece 
that showed a larger $\Delta C/T(T_{\rm N})$. 
In contrast, a larger $\Delta C/T(T_{\rm c}^{\rm h})$ appeared in a piece 
that showed a smaller $\Delta C/T(T_{\rm N})$. 
Moreover, 
the piece with the largest jump at $T_{\rm c}^{\rm l}$ had 
a small $\gamma_{\rm s}$ and the largest $RRR$, 
and these properties appeared in both heat-treated and annealed pieces. 
Thus, SC$^{\rm l}$ was concluded to coexist with 
the AFM having $T_{\rm N}$ = 2.2 K in a regular lattice 
as non-centrosymmetricity. 
The volume fractions of SC$^{\rm l}$ and SC$^{\rm h}$ change with 
the state of AFM ordering 
and the defects in crystal structures.

\section*{Acknowledgments}
We thank A. Sumiyama, T. Kohara, K. Ueda, and Y. Hasegawa for helpful discussions. 
This work was partially supported by a Grant-in-Aid for Scientific Research 
from the Ministry of Education, 
Culture, Sports, Science and Technology, Japan.


\begin{thebibliography}{99} 
\bibitem{EBauer} E. ~Bauer, G. ~Hilscher, H. ~Michor, Ch. ~Paul, 
                 E. ~W. ~Scheidt, A. ~Gribanov, Yu. ~Seropegin, 
                 H. ~No{\"e}l, M. ~Sigrist, and P. ~Rogl: 
                 Phys. Rev. Lett. {\bf 92} (2004) 027003. 
\bibitem{PAFrig} P. A. ~Frigeri, D. F. ~Agterberg, A. ~Koga, and M. ~Sigrist: 
                 Phys. Rev. Lett. {\bf 92} (2004) 097001. 
\bibitem{SFuji}  S. ~Fujimoto: 
                 J. Phys. Soc. Jpn. {\bf 75} (2006) 083704. 
\bibitem{NHaya}  N. ~Hayashi, K. ~Wakabayashi, P. A. ~Frigeri, and M. ~Sigrist: 
                 Phys. Rev. B. {\bf 73} (2006) 024504. 
\bibitem{HTana}  H. ~Tanaka, H. ~Kaneyasu, and Y. ~Hasegawa: 
                 J. Phys. Soc. Jpn. {\bf 76} (2007) 024715. 
\bibitem{YYana}  Y. ~Yanase and M. ~Sigrist: 
                 J. Phys. Soc. Jpn. {\bf 76} (2007) 043712. 
\bibitem{TTake}  T. ~Takeuchi, T. ~Yasuda, M. ~Tsujino, H. ~Shishido, 
                 R. ~Settai H. ~Harima, and Y. ~{\=O}nuki: 
                 J. Phys. Soc. Jpn. {\bf 76} (2007) 014702. 
\bibitem{ESchei} E-W. ~Scheidt, F. ~Mayr, G. ~Eickerling, P. ~Rogl 
                 and E. ~Bauer: 
                 J. Phys.: Condens. Matter {\bf 17} (2005) L121. 
\bibitem{KNaka}  K. ~Nakatsuji, A. ~Sumiyama, Y. ~Oda, T. ~Yasuda, R. ~Settai 
                 and Y. ~{\=O}nuki: 
                 J. Phys. Soc. Jpn. {\bf 75} (2006) 084717. 
\bibitem{YAoki}  Y. ~Aoki, A. ~Sumiyama, G. ~Motoyama, Y. ~Oda, 
                 T. ~Yasuda, R. ~Settai and Y. ~{\=O}nuki: 
                 J. Phys. Soc. Jpn. {\bf 76} (2007) 114708. 
\bibitem{KIzawa} K. ~Izawa, Y. ~Kasahara, Y. ~Matsuda, K. ~Behnia, T. ~Yasuda, 
                 R. ~Settai, and Y. ~{\=O}nuki: 
                 Phys. Rev. Lett. {\bf 94} (2005) 197002. 
\bibitem{MYogi}  M. ~Yogi, Y. ~Kitaoka, S. ~Hashimoto, T. ~Yasuda, R. ~Settai, 
                 T. D. ~Matsuda, Y. ~Haga, Y. ~{\=O}nuki, P. Rogl, and E. Bauer: 
                 Phys. Rev. Lett. {\bf 93} (2004) 027003. 
\bibitem{KUeda1} K. ~Ueda, K. ~Hamamoto, T. ~Kohara, G. ~Motoyama and Y. ~Oda: 
                 Physica B. \textbf{359-361} (2005) 374. 
\bibitem{KUeda2} K. ~Ueda, T. ~Kohara, G. ~Motoyama and Y. ~Oda: 
                 J. Mag. Mag. Mat. \textbf{310} (2007) 608. 
\bibitem{TYasu}  T. ~Yasuda, H. ~Shishido, T. ~Ueda, S. ~Hashimoto, 
                 R. ~Settai, T. ~Takeuchi, T. ~D. ~Matsuda, 
                 Y. ~Haga and Y. ~{\=O}nuki: 
                 J. Phys. Soc. Jpn. {\bf 73} (2004) 1657. 
\bibitem{NTate}  N. ~Tateiwa, Y. ~Haga, T. ~D. ~Matsuda, S. ~Ikeda, 
                 T. ~Yasuda, T. ~Takeuchi, R. ~Settai and Y. ~{\=O}nuki: 
                 J. Phys. Soc. Jpn. {\bf 74} (2005) 1903. 
\bibitem{Bauer2} E. ~Bauer, H. ~Kaldarar, A. ~Prokofiev, E. ~Royanian, 
                 A. ~Amato, J. ~Sereni, W. ~Br{\"a}mer-Escamilla, 
                 and I. ~Bonalde: 
                 J. Phys. Soc. Jpn. {\bf 76} (2007) 051009. 
\bibitem{GMoto1} G. ~Motoyama, S. ~Yamamoto, H. ~Takezoe, Y. ~Oda, 
                 K. ~Ueda and T. ~Kohara: 
                 J. Phys. Soc. Jpn. {\bf 75} (2006) 013706. 
\bibitem{GMoto2} G. ~Motoyama, M. ~Watanabe, K. ~Maeda, Y. ~Oda, 
                 K. ~Ueda and T. ~Kohara: 
                 J. Mag. Mag. Mat. \textbf{310} (2007) e126. 
\bibitem{NMetok} N. ~Metoki, K. ~Kaneko, T. D. ~Matsuda, A. ~Galatanu, T. ~Takeuchi, 
                 S. ~Hashimoto, T. ~Ueda, R. ~Settai, Y. ~{\=O}nuki and N. ~Bernhoeft: 
                 J. Phys.: Condens. Matter {\bf 16} (2004) L207. 
\bibitem{WHige}  W. ~Higemoto, Y. ~Haga, T. D. ~Matsuda, Y. ~{\=O}nuki, K. ~Ohishi, 
                 T. U. ~Ito, A. ~Koda, S. R. ~Saha and R. ~Kadono: 
                 J. Phys. Soc. Jpn. {\bf 75} (2006) 124713. 
\end{thebibliography}
\end{document}